\documentclass{jac}

\newcommand{\ket}[1]{| #1 \rangle}
\newcommand{\bra}[1]{\langle #1 |}
\newcommand{\cro}[1]{\left[#1\right]}
\newcommand{\braket}[2]{\langle #1 | #2 \rangle}
\newcommand{\incl}{\subseteq}
\newcommand{\trace}{\textrm{Tr}}

\newcommand{\Id}{\operatorname{Id}}

\newcommand{\Z}{\mathbb{Z}}
\newcommand{\joliA}{\mathcal{A}}
\newcommand{\joliB}{\mathcal{B}}
\newcommand{\joliC}{\mathcal{C}}
\newcommand{\joliH}{\mathcal{H}}

\newcommand{\joliL}{\mathcal{L}}

\newcommand{\joliN}{\mathcal{N}}
\newcommand{\pa}[1]{\left(#1\right)}
\newcommand{\acco}[1]{\left\{#1\right\}}

\newcommand{\XOR}{\operatorname{XOR}}

\begin{document}

\title{Quantization of Cellular Automata}

\author[Grenoble]{P. Arrighi}{Pablo Arrighi}

\address[Grenoble]{Universit\'e de Grenoble
\newline LIG, 46 Avenue F\'elix Viallet
\newline 38031 Grenoble Cedex FRANCE}
\email{pablo.arrighi@imag.fr}

\author[BS]{V. Nesme}{Vincent Nesme}
\address[BS]{Technische Universit\"at Braunschweig
\newline IMaPh, Mendelssohnstra\ss e 3
\newline 38106 Braunschweig DEUTSCHLAND}
\email{vincent.nesme@tu-bs.de} 

\begin{abstract}
Take a cellular automaton, consider that each configuration is a basis vector in some vector space, and linearize the global evolution function. If lucky, the result could actually make sense physically, as a valid quantum evolution; but does it make sense as a quantum cellular automaton? That is the main question we address in this paper. In every model with discrete time and space, two things are required in order to qualify as a cellular automaton: invariance by translation and locality. We prove that this locality condition is so restrictive in the quantum case that every quantum cellular automaton constructed in this way --- i.e., by linearization of a classical one --- must be reversible. We also discuss some subtleties about the extent of nonlocality that can be encountered in the one-dimensional case; we show that, even when the quantized version is non local, still, under some conditions, we may be unable to use this nonlocality to transmit information nonlocally. 
\end{abstract}


\maketitle

\section*{Introduction}

After some tries \cite{Watrous,Durr1,Durr2,Arrighi} at defining and studying quantum cellular automata, it is now believed to be fairly well understood how reversible quantum cellular automata (RQCA) should be defined, and what their basic properties are. As with classical cellular automata (CA), there are two levels on which RQCA are defined: as local transition functions or as global evolutions. The definition of RQCA proposed in \cite{Werner} focuses only on the properties of the global evolution, based on the two essential points of invariance by translation and locality. It was also proved in the same paper that each reversible cellular automaton could be ``quantized'' in a natural way, and the result would be a RQCA.
Furthermore, it was proved that RQCA can be implemented with local means, thereby reinforcing the parallel with CA; this was first done in the one-dimensional case \cite{Werner,ANW1}, the result being later extended to the general case \cite{ANW2}. Also, they involve no measurement procedure; the global evolution of a RQCA can be described by a unitary operator, while its decomposition as layers of local operations consists only of small unitary transformations.

We would like now to extend this framework to include cases where the global evolution is no longer described by a unitary operator, but by an isometry. This would be the first step in the investigation of nonreversible quantum cellular automata (NRQCA). The main problem with this topic, nowadays, is that there is no practical definition for such things. Our aim is to provide such a definition and work out the basic properties of NRQCA. 
Invariance by translation and locality as defined in \cite{ANW1} are still properties that NRQCA should obviously have.
In this paper we will ask and answer this question: when does the quantization of a CA have these properties? Since the translational invariance comes freely, the real question is: when is the quantization local?

Section~\ref{definitions} will be devoted to the mandatory definitions. We will be quick as we assume the reader is familiar with the basics of CA and somewhat familiar with quantum computing. We then show with theorem~\ref{thm_central} that the locality --- more precisely, the \emph{uniform} locality, cf. definition~\ref{locality} --- of the quantization is equivalent to reversibility, therefore extending the results presented in~\cite{Werner,ANW1}, and proving that no NRQCA can be constructed in this way. In Section~\ref{annexe}, we discuss the one-dimensional case. We show that, in some cases, even if the quantization is not uniformly local, it can still be local in a weaker sense which forbids some kinds of long-distance communications.

\section{Definitions}\label{definitions}

We will now introduce the basic definitions of quantum cellular automata.
For technical reasons, we will work mainly with finite configurations. This is because they are countable, as opposed to infinite configurations, and we want to have vector spaces of countable dimension, so as to simplify the formalism of \cite{Werner}. This distinction between finite and infinite configurations is not so important, as was shown in~\cite{ANW1}; anyway, we are only interested in locality conditions for quantizations of CA. We do not restrict the dimension of the space, which will be some positive integer $d$. We denote $q$ the quiescent state, and $\Sigma$ the rest of the alphabet, assuming $q\notin\Sigma$; the union of $\Sigma$ and $\acco{q}$ is denoted $q\Sigma$. The sets of finite configurations is denoted $\joliC_f$; it contains the elements of $\pa{q\Sigma}^{\Z^d}$ that are equal to $q$ almost everywhere on $\Z^d$.

Whilst configurations hold the basic states of an 
entire line of cells, and hence denote the possible basic 
states of the entire QCA, the global state of a QCA may 
well turn out to be a superposition of these. The following 
definition works because $\mathcal{C}_f$ is a countably 
infinite set.

\begin{defi}[Superpositions of configurations]~\label{superp}\\ 
Let $\mathcal{H}_{\mathcal{C}_f}$ be the Hilbert space of configurations, defined as follows. To each finite configuration $c$ is associated a unit vector $\ket{c}$, such that the family $\pa{\ket{c}}_{c\in\joliC_f}$ is an orthonormal basis of $\joliH_{\joliC_f}$. A \emph{superposition of 
configurations} is then a unit vector in $\joliH_{\joliC_f}$.
\end{defi}

We used here Dirac notation. Likewise, $\bra{c}$ denotes the dual of $\ket{c}$, i.e. the linear form on $\joliH_{\joliC_f}$ such that for all $d\in\joliC_f$, $\bra{c}\pa{\ket{d}}$, which is noted $\braket{c}{d}$, is equal to $\delta_{cd}$. These notations may then be combined the other way around, $\ket{c}\bra{c'}$ being the linear transformation of $\joliH_{\joliC_f}$ such that $\ket{c}\bra{c'}\ket{d}$ is, quite naturally, equal to $\braket{c'}{d}\ket{c}$.

\emph{States} on $\mathcal{H}_{\mathcal{C}_f}$ are nonnegative hermitian operators of trace $1$. For instance, for each superposition of configurations $\ket{\psi}$, $\ket{\psi}\bra{\psi}$ is a state, called in this case a \emph{pure} state.
Physically, states describe the actual state of matter; they bear all the information that can be measured in the system. The cells of our CA are in the pure state $\ket{\psi}\bra{\psi}$ when what lies on them is, with certainty, the superposition $\ket{\psi}$.
Actually, each state can be approximated by convex combinations of pure states. It means that the actual physical state of our CA at some moment can be described as a (possibly infinite) sum $\sum\limits_i p_i\ket{\psi_i}\bra{\psi_i}$, where the $p_i$'s are positive, $\sum\limits_i p_i =1$ and the $\ket{\psi_i}$'s are pairwise orthogonal.

We will be manipulating isometries a lot. Unitary operators should be well-known, but isometries are probably somewhat less familiar, so let us write down their definition. A linear operator $G:\joliH_{\joliC_f}\longrightarrow\joliH_{\joliC_f}$ is \emph{isometric} if and only if $\{G\ket{c}\,|\,c\in\mathcal{C}_f\}$ is an orthonormal family of $\mathcal{H}_{\mathcal{C}_f}$. This can also be expressed simply using the adjoint $G^\dagger$ of $G$. By definition, when $G$ is a endomorphism of $\joliH_{\joliC_f}$, $G^\dagger$ is the endomorphism of $\joliH_{\joliC_f}$ such that for every $\ket{\varphi},\ket{\psi}\in\joliH_{\joliC_f}$, $\bra{\varphi}G\ket{\psi}=\bra{\psi}G^\dagger\ket{\varphi}$. This way, $G^\dagger$ is indeed always unique; however, it is defined if and only if $G$ is continuous. When $G$ is isometric, $G$ is of course continuous, and actually, $G$ is isometric iff $G^\dagger G=\Id_{\joliC_f}$. If, moreover, $G$ is onto, it is said to be $\emph{unitary}$; so $G$ is unitary if and only if $G^\dagger G=GG^\dagger=\Id_{\joliC_f}$.

Now, we are talking about CA, whose one important feature is shift-invariance. The definition of shift-invariance in a quantum context, with all these linearizations, is actually no more tedious than in the classical case; here it is.

\begin{defi}[Shift-invariance]~\label{shift-invariance}\\
Consider the shift operation which takes configuration $\ldots c_{i-1}c_ic_{i+1}\ldots$ to $\ldots c'_{i-1}c'_ic'_{i+1}\ldots$ where, for all $i$, $c'_i=c_{i+1}$. Let $\sigma:\mathcal{H}_{\mathcal{C}_f}\longrightarrow\mathcal{H}_{\mathcal{C}_f}$ be its linear extension. A linear operator $G:\mathcal{H}_{\mathcal{C}_f}\longrightarrow\mathcal{H}_{\mathcal{C}_f}$ is said to be 
\emph{shift invariant} if and only if $G\sigma=\sigma G$.
\end{defi}

The second important feature of CA is their locality. In the classical case, we know that the locality is equivalent to the continuity of the global evolution on infinite configurations. Unfortunately, there does not seem to be such a result in the quantum case; at least it is not obvious what the right topology on superpositions of configurations should be. Therefore, the definition of locality proposed in \cite{Werner} is more concrete. It explicitly states that to know the state of some region of the space after an iteration of the CA, you only need to know the state of a slightly larger region beforehand. In the classical case, you would trivially deduce from this property that the global evolution stems from a local transition rule. In the quantum case however, things are not so simple as entanglement suddenly comes into play, and when $G$ is unitary it turns out \cite{Werner, ANW1,ANW2} you need to keep things locally reversible. 


To give the actual definition of locality, we first need to introduce some vocabulary. First, we will make abundant use throughout this paper of the Minkowski sum. For two subsets $\joliA$ and $\joliB$ of $\Z^d$, the Minkowski sum of $\joliA$ and $\joliB$, noted $\joliA+\joliB$, is the set $\acco{a+b/a\in \joliA, b\in \joliB}$. $\joliA-\joliB$ is naturally the Minkowski difference, $\acco{a-b/a\in \joliA, b\in \joliB}$.

$\joliH_{\joliC_f}$ has a natural structure of tensor product. Namely, for a subset $\joliA$ of $\Z^d$, let us note $\joliC_f\pa{\joliA}$ the set of the finite words on $\joliA$. Then $\joliH_{\joliC_f}$ is naturally isomorphic to $\joliH_{\joliC_f(\joliA)}\otimes\joliH_{\joliC_f\pa{\overline{\joliA}}}$, where $\overline{\joliA}$ denotes the complementary of $\joliA$ in $\Z^d$ and $\joliH_{\joliC_f(\joliA)}$ is the Hilbert space whose canonical basis is indexed by the elements of $\joliC_f(\joliA)$. That being said, there are two more definitions we need before moving on. The first one should be familiar, it is also known as ``trace out'' and occurs whenever a quantum system can be divided into two subsystems. Informally, if a system $S$ can be written as the tensor product of two subsystems $A$ and $B$, and given a state $\rho$ on $S$, you can chose to ignore completely what is going on $B$ and restrict your universe to $A$. The state you get on $A$ is then the restriction of $\rho$ to $A$.

\begin{defi}[Reduction]
Let $\rho$ be a state over $\joliH_{\joliC_f}$ and $\joliA$ a subset of $\Z^d$. One can write $\rho=\sum\limits_i \sigma_i\otimes\tau_i$, where the $\sigma_i$'s and $\tau_i$'s are respectively operators over $\joliH_{\joliC_f(\joliA)}$ and $\joliH_{\joliC_f\pa{\overline{\joliA}}}$. Then $\rho|_\joliA$, the \emph{reduction} of $\rho$ to $\joliA$, is a state on $\joliH_{\joliC_f(\joliA)}$ defined as $\sum\limits_i \trace\pa{\tau_i}\sigma_i$; this does not depend on the way $\rho$ was decomposed in the first place.\qed
\end{defi}

The following definition is the dual of the last one. Why it is its dual will appear in proposition~\ref{localization}.

\begin{defi}[Localization]
A linear endomorphism of $\joliH_{\joliC_f}$ is \emph{localized} in a subset $\joliA$ of $\Z^d$ if it is of the form $A\otimes\Id$, where $A$ is an endomorphism of $\joliH_{\joliC_f\pa{\joliA}}$ and $\Id$ is the identity on $\joliH_{\joliC_f\pa{\overline{\joliA}}}$.
\end{defi}

We can now explain the duality going on here with this lemma, which is stated and proved as lemma~3 in \cite{ANW1}.

\begin{lem}[Duality]\label{measure}~\\
Let $\joliH_0$ and $\joliH_1$ be Hilbert spaces, with $\joliH_0$ of finite dimension $p$. 
Let  $A, \rho, \rho'$ denote some elements of $\joliL(\joliH_0\otimes\joliH_1)$ with $\rho, \rho'$ having 
reductions $\rho|_0,\rho'|_0$ over $\joliH_0$.
We then have that $A$ is localized in $\joliH_0$ iff
\begin{center}
``for every states $\rho$ and $\rho'$, if $\rho|_0=\rho'|_0$ then $\trace(A\rho)=\trace(A\rho')$''.
\end{center}
Moreover we have that $\rho|_0=\rho'|_0$ is equivalent to 
\begin{center}
\hfill ``if $A$ is localized in $\joliH_0$, then $\trace(A\rho)=\trace(A\rho')$''.\qed
\end{center}
\end{lem}

The proposition \ref{localization} we introduce next comes from theorem~3 in \cite{ANW1}.
It entails structural reversibility, i.e. the fact that the inverse function of a RQCA is also a RQCA. Since we want now to talk about nonunitary operators, we have to restate it for general linear operators. We also have to extend the domain of localization of this operator from one cell to a set of cells.
It will also serve as a definition of locality for linear endomorphisms over $\joliH_{\joliC_f}$ --- which is not to be confused with localization. Note that the hypothesis of continuity for $G$ provides the existence of its adjoint $G^\dagger$.

It defines the locality ``at somewhere'' in the space. Intuitively, a global transition is said to be local at some locus if the physical state in this locus after the transition depends only on the physical state on a neighbourhood of this locus beforehand (this is point (i) of the proposition). Equivalently, one could say that the result of each measure done on this locus after the transition could be predicted beforehand by measures performed on its neighbourhood (that would be point (ii)).

\begin{prop}[Structural reversibility]~\\ \label{localization}
Let $G$ be a continuous linear endomorphism of $\joliH_{C_f}$, $\joliA$ and $\joliN$ respectively a subset and a finite subset of $\Z^d$. Suppose $G^\dagger$ 
 The two properties are equivalent:
\begin{itemize}
\item[(i)] For every states $\rho$ and $\rho'$, if $\rho|_{\joliA+\joliN}=\rho'|_{\joliA+\joliN}$ then $\pa{G\rho G^\dagger}|_\joliA=\pa{G\rho' G^\dagger}|_\joliA$.\\
\item[(ii)] For every operator $A$ localized in $\joliA$, $G^\dagger AG$ is localized in $\joliA+\joliN$.\\
\end{itemize}

When $G$ satisfies these properties, we say that $G$ is local at $\joliA$ with neighbourhood $\joliN$. If only $\joliA$ is given, we say that $G$ is local at $\joliA$ if there exists a finite subset $\joliN$ of $\Z^d$ such that $G$ is local at $\joliA$ with neighbourhood $\joliN$.

If $G$ is unitary, the following items are equivalent to (i) and (ii).

\begin{itemize}
\item[(iii)] For every states $\rho$ and $\rho'$ over the finite 
configurations, if $\rho|_{\joliA-\joliN}=\rho'|_{\joliA-\joliN}$ then $\pa{G^\dagger\rho 
G}|_\joliA=\pa{G^\dagger\rho' G}|_\joliA$.\\
\item[(iv)] For every operator $A$ localized in $\joliA$, then $G 
AG^\dagger$ is localized on the cells in $\joliA-\joliN$.
\end{itemize}
\end{prop}
\proof\ \\
$[(i)\Rightarrow(ii)]$. Suppose (i) and let $A$ be an operator 
acting on cell $0$. For every states $\rho$ and $\rho'$ such that 
$\rho|_\joliN=\rho'|_\joliN$, we have $\trace\pa{AG\rho 
G^\dagger}=\trace\pa{AG\rho' G^\dagger}$, using lemma \ref{measure} and our hypothesis that $\pa{G\rho G^\dagger}|_\joliA=\pa{G\rho' G^\dagger}|_\joliA$. We thus 
get $\trace\pa{G^\dagger AG\rho}=\trace\pa{G^\dagger AG\rho'}$. Since 
this is true of every $\rho$ and $\rho'$ such that 
$\rho|_{\joliA+\joliN}=\rho'|_{\joliA+\joliN}$, this means, again according to 
lemma \ref{measure}, that $G^\dagger AG$ is localized on the cells in $\joliA+\joliN$.\\
$[(ii)\Rightarrow (i)]$. Suppose (ii) and $\rho|_{\joliA+\joliN}=\rho'|_{\joliA+\joliN}$. 
Then, for every operator $B$ localized on the cells in $\joliN$, 
lemma \ref{measure} gives $\trace\pa{B\rho}=\trace\pa{B\rho'}$, 
so for every operator $A$ localized on cell 0, we get:
\begin{align*}
\trace\pa{AG\rho G^\dagger}&=\trace\pa{G^\dagger 
AG\rho}\\
&=\trace\pa{G^\dagger AG\rho'}\\
\trace\pa{AG\rho G^\dagger}&=\trace\pa{AG\rho'G^\dagger}
\end{align*}
Again by lemma \ref{measure}, this means $\pa{G\rho G^\dagger}|_0=\pa{G\rho' G^\dagger}|_0$.\\
Let us now assume $G$ is unitary.\\
$[(ii)\Rightarrow (iv)]$. Suppose $(ii)$ and let $A$ be an operator acting on cell $0$. Consider some operator $M$ acting on a cell $i$ which does not belong to $-\joliN$. According to our hypothesis we know that $G^{\dagger}MG$ does not act upon cell $0$, and hence it commutes with $A$. But $AB\mapsto GAG^{\dagger}GBG^{\dagger}=GABG^{\dagger}$ is a morphism, hence $GG^{\dagger}MGG^{\dagger}=M$ also commutes with $GAG^{\dagger}$. Because $M$ can be chosen amongst to full matrix algebra $M_d(\mathbb{C})$ of cell $i$, this entails that $GAG^{\dagger}$ must be the identity upon this cell. The same can be said of any cell outside $-\joliN$.\\
$[(iv)\Rightarrow (ii)]$, $[(iii)\Rightarrow (iv)]$, $[(iii)\Leftarrow (iv)]$ are symmetrical to 
$[(ii)\Rightarrow (iv)]$, $[(i)\Rightarrow (ii)]$, $[(ii)\Leftarrow (i)]$ just by interchanging the roles of $G$ and $G^{\dagger}$.\qed

We can now say that again in a mathematically rigorous way: a RQCA is a unitary operator on $\joliH_{\joliC_f}$ that is shift-invariant and local at the central cell. Indeed, in this case, the assumption of locality at the central cell implies the locality at each finite subset $\joliA$ of $\Z^d$. Moreover, this locality is uniform, in the sense that a same neihbourghood $\joliN$ can be chosen for all $\joliA$'s. However, if we remove this hypothesis of unitarity, things are not so simple and we have to make stronger hypotheses; hence the following definitions.

\begin{defi}[Locality]\label{locality}
A continuous linear endomorphism $G$ of $\joliH_{\joliC_f}$ is \emph{everywhere local} if, for every finite subset $\joliA$ of $\Z^d$, $G$ is local at $\joliA$. It is \emph{uniformly local} if there exists a finite subset $\joliN$ of $\Z^d$ such that for every finite subset $\joliA$ of $\Z^d$, $G$ is local at $\joliA$ with neighbourhood $\joliN$.
\end{defi}

\section{Linearization of Classical Automata}\label{plat_de_resistance}

Let $F:\joliC_f\to\joliC_f$ be a cellular automaton on finite configurations, $\widetilde{F}:\joliH_{\joliC_f}\to\joliH_{\joliC_f}$ its linearization. For it to have a physical meaning and earn its name of ``quantization'', it should be an isometry, i.e. $F$ should be one-to-one. We will nevertheless make a seemingly weaker assumption; we only assume $\widetilde{F}^\dagger$ to be defined, in order to be able to apply definition~\ref{locality} and ask when $\widetilde{F}$ is local; it turns out that this condition actually implies the injectivity of $F$.

In order for $\widetilde{F}^\dagger$ to be defined, we have to assume that $\widetilde{F}$ is continuous. Beware that this notion of continuity has nothing to do with any kind of topology on the set of words, and is therefore not related to the continuity of $F$, which is true by definition of a CA. For $\widetilde{F}$ to be continuous means that it is bounded on the unit sphere of $\joliH_{\joliC_f}$. This is equivalent to saying that $F$ is one-to-one. To verify this, let us first assume $F$ is one-to-one. Then $\widetilde{F}^\dagger$ is isometric, and consequently continuous. Let us now assume $F$ is not one-to-one. Since $F$ is defined on the finite configurations, for every $n$, there exists $x_n\in\joliC_f$ such that $x_n$ has a number of antecedents $\mu_n$ greater than $n$ --- just repeat as many times as needed some finite configuration having several antecedents. But then $\frac{1}{\sqrt{\mu_n}}\sum_{y\in\joliC_f/F(y)=x_n}\ket{y}$ is a unit vector whose image by $\widetilde{F}$, $\sqrt{\mu_n}\ket{x_n}$, has norm $\sqrt{\mu_n}$; hence, $\widetilde{F}$ is not bounded on the unit sphere, i.e. not continuous. To close this chapter on $\widetilde{F}^\dagger$, note that when it exists, it is defined as such:

$$\widetilde{F}^\dagger \ket{a}=\sum\limits_{u\in\joliC_f/F(u)=a}\ket{u}.$$

We therefore assume from now on that $F$ is one-to-one.
So, if you are given a word $w$ in the image of $F$, there is a unique $u\in\joliC_f$ such that $F(u)=w$. In general though, $u$ can not be computed locally from $w$. If it were possible to do that, the cellular automaton would be, by definition, reversible, and thus, according to \cite{Werner}, its linearization would be a \emph{bona fide} reversible quantum cellular automaton.

We will be monitoring $\XOR$ as an example, for which we will allow the quiescent state $q$ to be renamed $0$, the only letter in $\Sigma$ being $1$. $\XOR$ acts exactly as the usual $\XOR$: it sums modulo $2$ the bits in its neighbourhood $\acco{0;1}$. Of course, $\XOR$ is not reversible, since $11\ldots1$ and $00\ldots0$ are sent locally on the same word. It is one-to-one on finite configurations, though, while not surjective. It was already stated in proposition~1 of \cite{ANW1} that the quantization of nonreversible automata that are bijective on finite configurations could not be local, but that left the case of such automata as $\XOR$ unsettled. The following theorem does the job.

\begin{thm}\label{thm_central}
Suppose $F$ is one-to-one. Then $\widetilde{F}$ is uniformly local if and only if $F$ is reversible.
\end{thm}

\proof
Let us first briefly justify that when $F$ is reversible, $\widetilde{F}$ is uniformly local. This is essentially what states the lemma~4 of \cite{Werner}, though in this case it is the automaton as defined on infinite configurations that is quantized. It is quite straightforward to adapt the statement and the proof of this lemma to our formalism, to get the same result: if $F$ admits a neighbourhood $\joliN_C$ and an inverse neighbourhood $\joliN_I$, then $\joliN_C-\joliN_C+\joliN_I$ is a neighbourhood $\overline{\joliN}$; this is actually a direct consequence of lemma~\ref{gros_bordel}. However, there is a much
simpler proof that such a neighbourhood exists. First, decompose your automaton into block permutations, with auxiliary bits if needed. Linearize then each of these block permutations. The composition of all these local unitary transformations is then $\widetilde{F}\otimes\Id$, where $\Id$ is the identity on the auxiliary qubits, and the block decomposition from which it is constructed is a witness that $\widetilde{F}$ is uniformly local.

We now prove the other implication, in a way that can be seen as a generalization of the argument presented page~7 of \cite{ANW1}. It proceeds by contraposition, so let us first of all assume $F$ is not reversible. We will prove that for every set $\joliN$ there exists a set $\joliA$ such that $\widetilde{F}$ cannot satisfy the condition (i) of proposition~\ref{localization}; this will mean that $\widetilde{F}$ is not uniformly local. 

Let $\joliN$ be a finite subset of $\Z^d$. Since $F$ is not reversible, there exists a finite subset $\joliB$ of $\Z^d$ such that $F(x)|_{\joliB-\joliN}=F(y)|_{\joliB-\joliN}$ but $x|_{\joliB}\neq y|_{\joliB}$. Let $\joliA=\acco{s\in\Z^d/F(x)|_s\neq F(y)|_s}$; since $F(x)$ and $F(y)$ both are finite configurations, $\joliA$ is finite.

Let $\ket{\varphi_\pm}$ denote the superpositions of configurations $\frac{\ket{x}\pm\ket{y}}{\sqrt{2}}$, and let $\rho_\pm$ be the pure states $\ket{\varphi_\pm}\bra{\varphi_\pm}$. We are now going to prove that $\rho_+|_{\joliA+\joliN}=\rho_-|_{\joliA+\joliN}$, while $\left.\pa{\widetilde{F}\rho_+\widetilde{F}^\dagger}\right|_\joliA\neq \left.\pa{\widetilde{F}\rho_-\widetilde{F}^\dagger}\right|_\joliA$.

Since $F(x)$ and $F(y)$ are equal on $\joliB-\joliN$, $\joliA+\joliN$ does not intersect $\joliB$, so $x$ and $y$ differ on some point on the complement of $\joliA+\joliN$. Considering the partition of $\Z^d$ into $\joliA+\joliN$ and $\overline{\joliA+\joliN}$, we can thus write $\ket{x}=\ket{x_1}\otimes\ket{x_2}$ and $\ket{y}=\ket{y_1}\otimes\ket{y_2}$, where $x_1,y_1\in\joliC_f\pa{\joliA+\joliN}$, $x_2,y_2\in\joliC_f\pa{\overline{\joliA+\joliN}}$, and $x_2\neq y_2$. We then have 

\begin{align*}
\rho_\pm|_{\joliA+\joliN}&=\frac{1}{2}\pa{\ket{x}\bra{x}\pm\ket{x}\bra{y}\pm\ket{y}\bra{x}+\ket{y}\bra{y}}|_{\joliA+\joliN}
\\
&=\frac{1}{2}\left.\pa{\begin{array}{c}\ket{x_1}\bra{x_1}\otimes\ket{x_2}\bra{x_2} \pm\ket{x_1}\bra{y_1}\otimes\ket{x_2}\bra{y_2}\\\pm\ket{y_1}\bra{x_1}\otimes\ket{y_2}\bra{x_2}+\ket{y_1}\bra{y_1}\otimes\ket{y_2}\bra{y_2}\end{array}}\right|_{\joliA+\joliN}\\
\rho_\pm|_{\joliA+\joliN}&=\frac{1}{2}\pa{\ket{x_1}\bra{x_1}+\ket{y_1}\bra{y_1}}.\\
\end{align*}

Thus, the reductions of $\rho_+$ and $\rho_-$ and $\joliA+\joliN$ are indeed equal. Now, $\widetilde{F}\rho_\pm\widetilde{F}^\dagger=\ket{\psi_\pm}\bra{\psi_\pm}$, where $\ket{\psi_\pm}=\frac{\ket{F(x)}\pm\ket{F(y)}}{\sqrt{2}}$. Since $F(x)$ and $F(y)$ coincide on $\overline{\joliA}$, we actually have $\widetilde{F}\rho_\pm\widetilde{F}^\dagger=\sigma_1\otimes{\sigma_\pm}$, where $\sigma_1$ is a (pure) state over $\joliH_{\joliC_f\pa{\overline{\joliA}}}$, and the $\sigma_\pm$'s are states over $\joliH_{\joliC_f\pa{\joliA}}$. The reductions of $\widetilde{F}\rho_\pm\widetilde{F}^\dagger$ to $\joliA$ are then $\sigma_\pm$, which are distinct states since $\rho_+$ and $\rho_-$ where distinct to begin with.
\qed

Another way to present this proof is to appeal to the perennial Alice and Bob. We start with the state $\rho_+$. Alice and Bob have access to some cells of $\Z^d$, meaning that they can conjugate the state on $\joliH_{\joliC_f}$ with unitary operators, as long as these unitary operators are localized in the region of the space they were assigned. So let Alice and Bob's regions be respectively $\joliA$ and $\joliB$ as encountered in the proof of theorem~\ref{thm_central}. We will see how they can communicate through the use of $\widetilde{F}$, even though their regions could be at quite a large distance from each other, depending on $\joliN$.

Since $x|_\joliB\neq y|_\joliB$, Bob is able to transform at will $\rho_+$ into $\rho_-$, by performing a \emph{controlled phase-shift} on some cell where $x$ and $y$ differ. What that means informally is that, since Bob is able to tell the difference between $x$ and $y$ in his area, he can introduce a dissimetry between $\ket{x}$ and $\ket{y}$. Of course he could simply transform $\ket{y}$ by changing the letters of $y$ is some cells or something like that, but that would not allow him to communicate any faster than in the classical case. So what Bob does is to change $\ket{y}$ into $-\ket{y}$, something more immaterial, purely quantum and, in a way ``delocalized'', that will allow Alice to catch his message, which is one bit of information : ``did I or didn't I change $\rho_+$ into $\rho_-$?''. After Bob did his thing, $\widetilde{F}$ is applied to the state.

Now, Alice being able to actually read the message is due to the fact that her region contains all the cells where $F(x)$ and $F(y)$. As explained in the proof of theorem~\ref{thm_central}, the state after $\widetilde{F}$ has been applied is a tensor product of a state on $\joliA$ and a state on $\overline{\joliA}$, the state on $\overline{\joliA}$ not depending on the prior actions of Bob; therefore, the state on $\joliA$ does depend on them, so Alice must have a way to distinguish between them --- in this case she just has to perform a so-called \emph{swap-test}. Let us see for instance what happens with $\XOR$. Consider these two words in $\joliC_f$:

$$
\begin{array}{rcl}
x&=&\ldots 0000000000000000\ldots\\
y&=&\ldots 0011111111111100\ldots\\
\end{array}$$

Their images are

$$
\begin{array}{rcl}
F(x)&=&\ldots 0000000000000000\ldots\\
F(y)&=&\ldots 0100000000000100\ldots\\
\end{array}$$

Now put Bob on the middle of the stripe, and Alice at the two cells where the $1$'s are in $F(y)$. By following the protocol described in the proof of theorem~\ref{thm_central}, Bob can indeed send a bit of information to Alice. There is no doubt that this is a correct proof that $\widetilde{\XOR}$ is not uniformly local, but one might argue that this idea of an ``Alice'' surrounding Bob makes little sense: surely if Alice can be present at two faraway places in the stripe at the same time, it means he must have some way to go from one place to the other, and since in the middle stands Bob, why would she bother using $\widetilde{\XOR}$ to send her message? Cannot we find another protocol where Alice stands either on the left or on the right of Bob, but on only one side at a time? Actually, no, we cannot, and this is related to the fact that $\widetilde{\XOR}$, while not uniformly local, is still everywhere local: if Bob is forbidden the access to the cells located between Alice's positions, then he cannot transmit her any message. The proof of this assertion is the object of the next section.

\section{Everywhere Locality in the One-dimensional Case}\label{annexe}


The question is: when is the quantization of a one-to-one CA everywhere local? We are going now to give a proof that in the one-dimensional case, it is equivalent to the openness of $F_\infty$, the extension of $F$ to the set $\joliC_\infty$ of infinite configurations; so let us fix the dimension $d$ to $1$ for this section.

First, it might be useful to remind what it means for $F_\infty$ to be open. $\joliC_\infty$ comes with the usual topology; namely, a base of open sets is given by the sets $\acco{v\in\joliC_\infty/v_\joliA=w_\joliA}$, for $w\in\joliC_\infty$ and $\joliA$ a finite subset of $Z^d$. By definition, $F_\infty$ is \emph{open} if for every open subset $O$ of $\joliC_\infty$, $F_\infty(O)$ is open.

\begin{prop}\label{open}
$\widetilde{F}$ is everywhere local if and only if $F_\infty$ is open.
\end{prop}
\proof
We will appeal to \cite{Kurka}. According to its theorem~5.45, $F_\infty$ is open iff it is left and right-closing. The definitions of left and right-closingness may be found in definition~5.38. First, $x$ and $y$ in $\joliC_\infty$ are said to be \emph{left-asymptotic} (respectively \emph{right-asymptotic}) when there is some $n\in \Z$ such that for every $k<n$ (resp. $k>n$), $x_k=y_k$. 
By definition, $F_\infty$ is left-closing (respectively right-closing) if, for every $x,y\in\joliC_\infty$ that are left-asymptotic (resp. right-asymptotic), if $F(x)=F(y)$ then $x=y$. We now translate these conditions on de Bruijn diagrams.

Let us recall briefly what we mean by Bruijn diagrams. Let $n$ be an integer such that $\cro{-n;n+1}$ is a neighbourhood for $F$.
We note $F_0$ the function from $\pa{q\Sigma}^{\cro{-n;n+1}}$ to $q\Sigma$ which computes locally $F$ on cell $0$, from the knowledge of the stripe on $\cro{-n;n+1}$. Then the associated de Bruijn diagram is a graph whose vertices are indexed by the pairs $(u,v)\in q\Sigma^{\cro{-n;n}}\times q\Sigma^{\cro{-n;n}}$. There is an edge from $(u,v)$ to $(u',v')$ if and only if

\begin{itemize}
\item for $i\in[-n;n[$, $u_{i+1}=u'_{i}$ and $v_{i+1}=v'_{i}$
\item $F_0(u_{-n}u_{-n+1}\ldots u_n u'_n)=F_0(v_{-n}v_{-n+1}\ldots v_n v'_n)$.
\end{itemize}

The first thing we want to note is that the strongly connected component (SCC) of $(q,q)$ in the de Bruijn diagram includes the diagonal $\Delta$ of $q\Sigma^{\cro{-n;n}}\times q\Sigma^{\cro{-n;n}}$, i.e. the elements of the form $(u,u)$.

To each pair of words $(u,v)\in\joliC_{\infty}\times\joliC_{\infty}$ such that $F(u)=F(v)$ is associated a bi-infinite path on the de Bruijn diagram, and vice-versa. In this respect, we see that ``$F_\infty$ is left-closing'' is equivalent to ``every infinite path starting from $\Delta$ stays forever in $\Delta$'', while ``$F_\infty$ is left-closing'' is the dual statement that ``every bi-infinite path ending in $\Delta$ is completely included in $\Delta$''. Thus, $F_\infty$ is open iff there is no connection, in or out, between $\Delta$ and any cycle of the de Brujin diagram not included in $\Delta$.

Now, what does it mean on this diagram for $\widetilde{F}$ to be everywhere local? If we follow the proof of theorem~\ref{thm_central}, we see this means that there exists an integer $k$ such that for every integer $n$, if $F(x)$ is known on $\overline{[-n;n]}$, then $x\in\joliC_f$ is determined on $\overline{[-n-k;n+k]}$. On the de Bruijn diagram, it means that there exists an integer $k$ such that any path starting from $(q,q)$ must stay in $X$ until $k$ steps before the end, and that every path ending in $(q,q)$ must stay in $X$ after $k$ steps. This also means that $X$ is not connected to any cycle not included in $\Delta$. 

Suppose $F_\infty$ is not open. Without loss of generality, we assume there is a path from a cycle not included in $\Delta$ to $\Delta$. This cycle is given by two distinct finite words $v$ and $v'$ of same length such that $F(\ldots vvvv\ldots)=F(\ldots v'v'v'v'\ldots)$; the path from this cycle to $(q,q)$ is given by two words of same lenght $w$ and $w'$, such that $F(\ldots vvvwqqq\ldots)=F(\ldots v'v'v'w'qqq\ldots)$ . Let $\cro{-n;n}$ be a neighbourhood for $F$ and $k$ a positive integer. Now consider the finite configurations $x_k=\ldots qqqv^kwqqq\ldots$ and $y_k=\ldots qqqv'^kw'qqq\ldots$, where the first letter of the first $v$ has position 0. Almost everywhere, $(x_k,y_k)$ follows a path on the de Bruijn diagram. The only points where $(x_k,y_k)$ does not follow an edge of this diagram is at the transition between cells $-1$ and $0$. So $\joliA_k=\acco{i\in\Z/F(x_k)\neq F(y_k)}$ is included in $[-n-1;n]$, and does not depend on $k$ when $k$ is large enough; let's define $\joliA=\lim\limits_{k\to\infty}\joliA_k$. Let $\joliB_k\incl\Z$ be the singleton consisting of the rightmost cell where $x_k$ and $y_k$ differ. Since $v\neq v'$, its emplacement is at least $k-1$. Let $\joliN$ be a finite subset of $\Z$; for a large enough $k$, we have the following properties:

\begin{itemize}
\item $F(x_k)|_{\joliB_k-\joliN}=F(y_k)|_{\joliB_k-\joliN}$
\item $x_k|_{\joliB_k}\neq y_k|_{\joliB_k}$
\item $\joliA=\acco{i\in\Z/F(x_k)\neq F(y_k)}$.
\end{itemize}

Then, according to the proof of theorem~\ref{thm_central}, $\widetilde{F}$ is not local at $\joliA$ with neigbourhood $\joliN$. Since we showed that there exists $\joliA$ such that this is true for any $\joliN$, we have indeed just proven that $\widetilde{F}$ is not everywhere local.

Now, what remains to prove is that when $F_\infty$ is open, $\widetilde{F}$ is everywhere local. To do that we will strengthen a little bit the lemma~4 of \cite{Werner}. But first we need to explain a property of one-dimensional open automata. Suppose $F_\infty$ is open and let $\joliA$ be a finite subset of $\Z$ and $x$ and $y$ two words such that $F(x)|_{\overline{\joliA}}=F(y)|_{\overline{\joliA}}$. Say $\joliA$ is included in $\cro{-n;n}$, $\cro{-k;k}$ is a neighbourhood for $F$ and $l$ is the number of vertices in the de Bruijn diagram.
If we look at $(x,y)$ as a run in this diagram, then we follow edges except perhaps in $\cro{-n-k;n+k}$. But since there are no loops connected in one way or another do $\Delta$, and we have to join $\Delta$ at $\pm\infty$, this means we are always in $\Delta$ except perhaps in $\cro{-n-k-l;n+k+l}$, to give a rough bound. So there exists a finite subset $\joliN_I$ of $\Z$, which does not depend on $x$ nor $y$ --- though it may depend on $\joliA$ --- such that $x|_{\overline{\joliA+\joliN_I}}=y|_{\overline{\joliA+\joliN_I}}$. Now all is needed to complete the proof is the next (and last) lemma, which, as announced, is but a gentle strengthening of the lemma~4 of \cite{Werner}.

\begin{lem}\label{gros_bordel}
Let $F$ be a one-to-one automaton with neighbourhood $\joliN_C$. Let $\joliA$ and $\joliN_I$ be finite subsets of $\Z$ such that for all $x,y\in \joliC_f\pa{\Z}$, if $F(x)|_{\overline{\joliA}}=F(y)|_{\overline{\joliA}}$, then $x|_{\overline{\joliA+\joliN_I}}=y|_{\overline{\joliA+\joliN_I}}$. Suppose $\joliN_C$ and $\joliN_I$ contain $0$. Then $\widetilde{F}$ is local at $\joliA$ with neighbourhood $\joliN=\joliN_C-\joliN_C+\joliN_I$.
\end{lem}
\proof
Let $\joliA\incl \Z^d$. Let $\rho$ and $\rho'$ be states over $\joliH_{\joliC_f}$ such that $\rho|_{\joliA+\joliN}=\rho'|_{\joliA+\joliN}$. We have to prove $\pa{\widetilde{F}\rho\widetilde{F}^\dagger}|_{\joliA}=\pa{\widetilde{F}\rho'\widetilde{F}^\dagger}|_{\joliA}$.

Let us write $\rho=\sum\limits_{a,b\in\joliC_f}\lambda_{a,b}\ket{a}\bra{b}$ and $\rho'=\sum\limits_{a,b\in\joliC_f}\lambda'_{a,b}\ket{a}\bra{b}$. Then 

$$\rho|_{\joliA+\joliN}=\sum\limits_{a,b/a_{\overline{\joliA+\joliN}}=b_{\overline{\joliA+\joliN}}}\lambda_{a,b}\ket{a_{\joliA+\joliN}}\bra{b_{\joliA+\joliN}}=\sum\limits_{x,y\in A^{\joliA+\joliN}} \pa{\sum\limits_{u\in A^{\overline{\joliA+\joliN}}}  \lambda_{x.u,y.u}} \ket{x}\bra{y}.$$

Ergo, the hypothesis $\rho|_{\joliA+\joliN}=\rho'|_{\joliA+\joliN}$ may be translated as

$$\forall x,y\in A^{\joliA+\joliN}\quad \sum\limits_{u\in A^{\overline{\joliA+\joliN}}}  \lambda_{x.u,y.u}=\sum\limits_{u\in A^{\overline{\joliA+\joliN}}}  \lambda'_{x.u,y.u}.$$

For $x,y\in A^{\joliA+\joliN}$, let $\alpha(x,y)$ be the set of couples $(a,b)$ of words in $\joliC_f$ such that $a_{\joliA+\joliN}=x$, $b_{\joliA+\joliN}=y$ and $a_{\overline{\joliA+\joliN}}=b_{\overline{\joliA+\joliN}}$. Then the hypothesis is equivalent to

\begin{equation}
\label{hypothèse}
\forall x,y\in A^{\joliA+\joliN} \sum_{(a,b)\in\alpha(x,y)}\lambda_{a,b}= \sum_{(a,b)\in\alpha(x,y)}\lambda'_{a,b}.	
\end{equation}

Let us now try translating our aim in the same way. First we have 

$$\widetilde{F}\rho\widetilde{F}^\dagger=\sum\limits_{a,b\in\joliC_f}\lambda_{a,b}\ket{F(a)}\bra{F(b)}=\sum\limits_{c,d\in F\pa{\joliC_f}}\lambda_{F^{-1}(c),F^{-1}(d)}\ket{c}\bra{d}$$

$$\pa{\widetilde{F}\rho\widetilde{F}^\dagger}|_{\joliA}=\sum\limits_{c,d\in F\pa{\joliC_f}/c_{\overline{\joliA}}=d_{\overline{\joliA}}}\lambda_{F^{-1}(c),F^{-1}(d)}\ket{c_{\joliA}}\bra{d_{\joliA}}$$

$$\pa{\widetilde{F}\rho\widetilde{F}^\dagger}|_{\joliA}=\sum\limits_{z,t\in A^{\joliA}} \pa{\sum\limits_{u\in A^{\overline{\joliA}}}  \lambda_{F^{-1}\pa{z.w},F^{-1}\pa{t.w}}} \ket{z}\bra{t}.$$

So what we want to prove is that, for every $z$ and $t$ in $A^\joliA$,

$$\sum\limits_{w\in A^{\overline{\joliA}}}  \lambda_{F^{-1}\pa{z.w},F^{-1}\pa{t.w}}=\sum\limits_{w\in A^{\overline{\joliA}}}  \lambda'_{F^{-1}\pa{z.w},F^{-1}\pa{t.w}},$$

with the convention that these numbers are $0$ when $F^{-1}$ is not appliable. For $z,t\in A^{\joliA}$, let $\beta(z,t)$ be the set of couples $(a,b)$ of words in $\joliC_f$ such that $F(a)_{\joliA}=z$, $F(b)_{\joliA}=t$ and $F(a)_{\overline{\joliA}}=F(b)_{\overline{\joliA}}$.
What we want to prove from (\ref{hypothèse}) is the equivalent to

\begin{equation}
\forall z,t\in A^{\joliA} \sum\limits_{(a,b)\in\beta(z,t)}\lambda_{a,b}= \sum\limits_{(a,b)\in\beta(z,t)}\lambda'_{a,b}.
\end{equation}

We will prove this by showing that for each $z,t\in A^\joliA$, there is some set $\gamma(z,t)$ such that $\beta(z,t)=\coprod\limits_{(x,y)\in\gamma(z,t)}\alpha(x,y)$, ie
$\beta(z,t)$ is the disjoint union of the $\alpha(x,y)$'s for $(x,y)$ in $\gamma(z,t)$.

On the one hand, it is quite immediate by definition that, when $(x,y)\neq (x',y')$, $\alpha(x,y)$ and $\alpha(x',y')$ are disjoint. On the other hand, by hypothesis, every $(a,b)$ of $\beta(z,t)$ is in some $\alpha(x,y)$, so that $\gamma(z,t)$ may be found in this simple way: 
for each $(a,b)$ in $\beta(z,t)$, find the unique $\pa{x_{a,b},y_{a,b}}$ such that $(a,b)$ is in $\alpha\pa{x_{a,b},y_{a,b}}$, and then define $\gamma(z,t)$ to be the set of all these $(x,y)$'s you found. The only problem is that you could add unwanted $(a,b)$'s by doing so; we need only checking that this is not the case. In other words, we have to prove that whenever the intersection between $\alpha(x,y)$ and $\beta(z,t)$ is nonempty, then the former is included in the latter.

So, let $(a,b)$ be an en element of $\alpha(x,y)\cap\beta(z,t)$ and $(a',b')$ an other element of $\alpha(x,y)$. First of all, since $a$ and $a'$ coincide on $\joliA+\joliN$ (where they are equal to $x$), and in particular on $\joliA+\joliN_C$, then $f(a)$ and $f(a')$ coincide on $\joliA$, thus $F(a')_\joliA=F(a)_\joliA=z$. Likewise, of course, $F(b')_\joliA=t$.

Then, by hypothesis and since $\joliA$ is finite and $F(a)_{\overline{\joliA}}=F(b)_{\overline{\joliA}}$, $a$ and $b$ coincide on $\overline{\joliA+\joliN_I}$, not only on $\overline{\joliA+\joliN}$. This implies that $x$ and $y$ must coincide on $\pa{\joliA+\joliN}\cap\overline{\joliA+\joliN_I}$, and as a consequence $a'$ and $b'$ do also coincide on $\overline{\joliA+\joliN_I}$; thus $F(a')$ and $F(b')$ coincide on $\overline{\joliA+\joliN_I-\joliN_C}$.

Lastly, since $a$ and $a'$ coincide on $\joliA+\joliN=\joliA+\joliN_C-\joliN_C+\joliN_I$, so do $F(a)$ and $F(a')$ on $\joliA-\joliN_C+\joliN_I$. Likewise, $F(b)$ and $F(b')$ coincide on that same interval. However, $F(a)$ and $F(b)$ coincide on $\overline{\joliA}$, by hypothesis; ergo, $F(a')$ and $F(b')$ coincide on $\overline{\joliA}\cap \pa{\joliA-\joliN_C+\joliN_I}$. Put it together, you finally get that $F(a')$ and $F(b')$ coincide on $\overline{\joliA}$; Q.E.D.\qed

$\XOR_\infty$ being, as can be checked easily on its de Bruijn diagram, open, it is thus everywhere local, which also means Alice has to surround Bob in order to receive his long-distance calls. On the contrary, the modified version of $\XOR$ that was defined in the definition~11 of \cite{ANW1} is not open on the infinite configurations, which is why we were able to find a protocol where Bob and Alice lie on two distinct sides of the stripe.

\section{Conclusion}

Starting only with the assumption that we should be able to use the adjoint of $\widetilde{F}$, this implied it should be isometric, thus convey a physical meaning as a valid quantum evolution. If we then add the constraint that it should be uniformly local --- something that you would certainly expect a cellular automaton to verify in any model --- it turns out $F$ has to be reversible, so that $\widetilde{F}$ is part of the already well-known class of RQCA. This is good news in a way: the notion of a RQCA is a robust one; however, it could nevertheless be considered a downside.
Indeed, as stated in the introduction, RQCA are now believed to be fairly well understood, so the next challenge is understanding nonreversible quantum cellular automata. It would certainly have been of great help to be able to construct such NRQCA by quantizing nonreversible CA. Alas, this paper shows that such a thing is impossible. Quantizing one-dimensional open non-reversible automata certainly provides puzzling entities, but no quantum CA; there remains however an interesting open question about the generalization of proposition~\ref{open} to higher dimensions.

Then again, the most important question right now is: what are NRQCA? Can they be defined from their global evolution in a reasonably simple way? This question, in its most general form, includes the same one concerning randomized automata instead of quantum ones, since classical randomness is part of the quantum world, and as far as we know this question has been little studied. Let us ask it in a more precise way: what is the property on the global evolution of probability distributions that characterizes randomized cellular automata, i.e. those transformations that can be written as a finite number of layers, each of them consisting of a tiling of identical blocks performing some local random transformation?

\section*{Acknowledgements}
We would like to thank Guillaume Theyssier for pointing out useful references and Reinhard Werner for asking the right questions and providing useful advices and encouragement. Also, a special thanks to the reviewer who remarked that a proof was absent and another one unclear, two remarks leading to the discovery of two mistakes in the first version of the paper, which led in turn to substantial rewriting. Our best hope is not to have introduced too many new mistakes in the process.

\end{document}